\documentclass[aps,prb,twocolumn,superscriptaddress]{revtex4}
\usepackage{psfrag}
\usepackage{graphicx}
\usepackage{psfig}
\usepackage{subfigure}
\usepackage{moreverb}
\usepackage{amsbsy}
\usepackage{amsmath}
\usepackage{amssymb}   
\newcommand{\aaa}    {{\rm \bf A }}

\newcommand{\hh}     {{\rm \bf h }}

\newcommand{\uu}     {{\rm \bf u }}
\newcommand{\MM}     {{\rm \bf M }}
\newcommand{\kk}     {{\rm \bf k }}
\newcommand{\WW}     {{\boldsymbol{\Omega}}}

\newcommand{\xx}     {{\rm \bf x}}

\newcommand{\fin}    {f_i^{(n)}}
\newcommand{\fnij}   {f_{ij}^{(n)}}

\newcommand{\fjn}    {f_j^{(n)}}
\newcommand{\ffn}    {{\rm \bf f}^{(n)}}

\newcommand{\qin}    {q^{(n)}}
\newcommand{\cin}    {c^{(n)}}

\newcommand{\df}     {\equiv}
\newcommand{\ee}     {{\rm e}}

\newcommand{\cross}  {\times}
\newcommand{\diff}   {{\rm\, d}}
\newcommand{\ceq}[1] {(\ref{#1})}

\newcommand{\beq}    {\begin{equation}}
\newcommand{\enq}    {\end{equation}}
\newcommand{\lag}  {\mathcal L}
\newcommand{\intvm}  {\int_{V_M}}
\newcommand{\intvr}  {\int_{V}}
\newcommand{\Heff}   {H_{\rm eff}}
\newcommand{\HHeff}  {{\rm \bf H}_{\rm eff}}
\newcommand{\Nint}   {{\mathbb{N} }}
\begin{document}

\title{Dynamics of magnetization coupled to a thermal bath of elastic modes}

\author{Enrico Rossi}
\affiliation{Department of Physics, University of Texas at Austin, Austin TX 78712}
\author{Olle G. Heinonen}
\affiliation{Seagate Technology, 7801 Computer Ave. South, Bloomington, MN 55435}
\author{A.H. MacDonald}
\affiliation{Department of Physics, University of Texas at Austin, Austin TX 78712}
\date{today}

%
%
\begin{abstract}
 We study the dynamics of magnetization coupled to a thermal bath of elastic modes
 using a system plus reservoir approach with realistic
 magnetoelastic coupling.  After integrating out the elastic modes
 we obtain a self-contained equation for the dynamics of the magnetization. 
 We find explicit expressions
 for the memory friction kernel and hence, {\em via} the Fluctuation-Dissipation 
 Theorem, for the spectral density of the magnetization thermal fluctuations.
 For magnetic samples in which the single domain approximation
 is valid, we derive an equation for the dynamics of the uniform mode.
 Finally we apply this equation to study the dynamics of the
 uniform magnetization mode in insulating ferromagnetic thin films. 
 As experimental consequences we find that 
 the fluctuation correlation time is of the order of
 the ratio between the film thickness, $h$, and the speed of sound
 in the magnet and that the line-width of the
 ferromagnetic resonance peak should scale as $B_1^2h$
 where $B_1$ is the magnetoelastic coupling constant.
\end{abstract}
%
%

\maketitle

%
%
\section{Introduction}
Thermally induced fluctuations of the magnetization
are responsible for one fundamental limit on the signal to noise
ratio of small magnetoresistive sensors \cite{smith_apl_01}.
The noise scales inversely with the volume
of the sensors and peaks at frequencies \cite{olle_immm_1,olle_immm_2}
that are now close to the ever increasing data rate
of magnetic storage devices. The increase of data
rates combined with the continuing decrease of the dimensions
of the sensors makes magnetic noise inevitable and motivates work aimed
at achieving a detailed understanding of its character.

The standard approach toward modeling of magnetization fluctuations
is to start from the Landau-Lifshitz-Gilbert-Brown equation \cite{brown_pr_63}
\beq
 \frac{\partial \WW}{\partial t} = 
  \frac{\gamma}{M_s}\WW\cross
  \left[
        \frac{\delta E}{\delta\WW} +
        {\bf h}
  \right] + 
  \alpha\WW\cross\frac{\partial\WW}{\partial t},
 \label{eq_llg}
\enq
where $\gamma$ is the gyromagnetic ratio,
$
\WW = \MM/M_s
$ is the magnetization direction,
$\MM$ is the magnetization, $M_S$ the magnitude of the saturation magnetization,
$E$ the free energy and ${\bf h}$ a random magnetic field.  This equation assumes that 
the characteristic time scale of the magnetization dynamics is longer than the typical
time scale of the environment that is responsible for the dissipative term proportional
to $\alpha$.  In practice the use of this equation is partially inconsistent, resulting in some 
practical limitations to its application \cite{garanin_prb_97,smith_jap_01}. 
The source of the problem is that the dissipation is  local in time.
Because of the fluctuations dissipation theorem,
this implicitly requires the random field to have white noise properties i.e.
to have zero autocorrelation time.
Since the contribution of the random field to the magnetization dynamics 
$\gamma\WW\cross {\bf h}$ depends on $\WW$, equation \ceq{eq_llg}
exhibits white multiplicative noise \cite{palacios_prb_98}.
It follows that in order to integrate
equation\ceq{eq_llg} reliably we need to track the
evolution of $\WW$ on very short time scales 
for which the white noise approximation for ${\bf h}$ is
likely to be unphysical.

In this paper we address the physics that determines the correlation time of the random field.
We start in section II by considering a formal model of a magnetic system coupled to an 
environment and specialize in Section III to an environment consisting of elastic modes.
In Section IV we consider the case in which a single magnetic mode
corresponding to 
coherent evolution of the magnetization in a small single-domain system
is coupled to the elastic environment.  In Section V we consider a thin film geometry in 
which the magnetization is coupled to elastic modes of the system and its substrate. 
Finally in Section VI we conclude with a discussion of the possible role of other 
sources of dissipation, in particular dissipation due to particle-hole excitations in 
the case of metallic ferromagnets.
 
%
\section{Generic reservoir}
Calling ${q_n}$ the degrees of freedom of the reservoir, we 
consider the following form for the total Lagrangian:
\begin{equation}
 \lag = \lag_S[\WW(\xx),\dot\WW(\xx)] +
        \lag_R[q_n,\dot q_n] +
	\lag_I[\WW(\xx),q_n]      - 
        \Delta \lag[\WW(\xx)],
 \label{eq_lag_tot}
\end{equation}
where $\lag_S[\WW(\xx),\dot\WW(\xx)]$ is the Lagrangian that describes
the dynamics of the magnetization when not coupled to
external degrees of freedom,
$\lag_R[q_n,\dot q_n]$ is the Lagrangian for the reservoir and
$\lag_I[\WW(\xx),q_n]$ is the interaction Lagrangian that couples
the magnetization to the reservoir degrees of freedom.
The term $\Delta \lag[\WW(\xx)]$ is a counter term that depends on $\WW$ and  the
parameters of the reservoir but not on the dynamic variables of the reservoir 
\cite{cl,weiss_book}. 
This term is introduced to compensate a renormalization of the energy
of the system caused by its coupling to the reservoir \cite{cl}.

The Landau-Liftshitz equations for the decoupled system magnetization follow from the
magnetic Lagrangian, 
\beq
 \lag_S = \int_{V_M}\left[\frac{M_s}{\gamma}\aaa[\WW]\cdot\dot\WW - E_s[\WW]\right]\diff\xx , 
 \label{eq_lag_s}
\enq
where $\aaa$ is a vector field defined by the equation:
$
 \nabla_{\WW}\cross\aaa[\WW] = \WW
$
and
$E_S[\WW]$ is the magnetic free energy functional 
and $V_M$ the volume of the ferromagnet.
We model the reservoir as a set of classical degrees of freedom:
\beq
 \lag_R = \frac{1}{2}\sum_n m_n\dot q_n^2 - E_R(q_n).
 \label{eq_lag_r}
\enq
The Euler-Lagrange equations for the total Lagrangian \ceq{eq_lag_tot} 
yield the following coupled dynamical equations:
\begin{align}
  m_n\ddot q_n =& \frac{\partial}{\partial q_n}[\lag_R(q_n,\dot q_n) + \lag_I[\WW,q_n]] 
  \label{eq_dyn_qn_gen}\\
       \frac{\partial\WW}{\partial t} =& \WW\cross
       \frac{\gamma}{M_s}
       \frac{\delta}{\delta\WW}[E_S[\WW,\dot\WW] - \lag_I[\WW, q_n] + \Delta\lag[\WW]].
  \label{eq_dyn_ww_gen}
\end{align}
When $\lag_I$ is linear in the coordinates of the bath,
we can formally integrate \ceq{eq_dyn_qn_gen} to get
$\qin(t)$ as a function only of the initial conditions and
$\WW$ and then insert the result in \ceq{eq_dyn_ww_gen} to
eliminate the reservoir coordinates from the dynamical equations
for $\WW$, integrating out the reservoir degrees of freedom.
An example of the application of this procedure
for a quantum mechanical model of the interaction
between magnetization and reservoir degrees of
freedom can be found in Ref. \onlinecite{rebei_prb_03}.

%
\section{Magnetization coupled to elastic modes: general}

If we consider only long wavelength vibrations
we can treat the lattice as a continuous medium and use
results from elasticity theory. 
The potential energy functional, $E_R$,
of the elastic medium can then be expressed in terms of the  strain tensor $u_{i,j}$,
$$
u_{ij}\df\frac{1}{2}\left(\frac{\partial u_i}{\partial x_j} + 
                          \frac{\partial u_j}{\partial x_i}
                    \right) 
$$
where ${\bf u}$ is the  displacement vector field. 

We want to study the dynamics of the magnetization when coupled
to elastic deformations of the system \cite{suhl_ieee}.
We will be interested in applying our results to polycrystalline 
elastic media which can be treated as isotropic to a good approximation. 
(It's quite straightforward, albeit quite tedious,
to extend our results to the case of non-isotropic media with specific
lattice symmetries). For isotropic elastic media it follows from
general symmetry considerations that, to lowest
order, we can express the magnetoelastic energy in the form, \cite{kittel_rmp_49}
\beq
  E_I = B_1\sum_{i,j=1}^3\int_{V_M} \Omega_i \Omega_j u_{ij}\diff \xx 
  \label{eq_def_ei}
\enq
where $B_1$ is the magnetoelastic coupling constant.
For the case of soft ferromagnet thin films, the main contribution to the magnetoelastic
energy will be given by the magnetostatic energy dependence
on strain. This contribution to $E_I$ is normally referred
as the the form effect \cite{lee55}.
The constant $B_1$ can be extracted from magnetostriction data. For an isotropic elastic medium with
isotropic magnetostriction, $\lambda$, we have \cite{kittel_rmp_49} that 
\beq
 B_1 = \frac{3}{2}\lambda\frac{E}{2-\sigma},
 \label{eq_b1}
\enq
where $E$ is the Young's modulus and $\sigma$ the 
Poisson's ratio.

The Lagrangian for an elastic reservoir
$\lag_R$ is,
\beq
 \lag_R = \frac{1}{2}\intvr\rho\; \dot\uu^2\diff\xx - E_R,
\enq
where $\rho$ is the mass density, $V$ the total volume of
the elastic medium (magnetic film plus substrate) 
and $E_R$  is given by \cite{landau_mech}:
\beq
   E_R =\intvr\left[\frac{E}{2(1+\sigma)}\sum_{i,j=1}^3 u_{ij}^2 +
                   \frac{\sigma E}{2(1+\sigma)(1-2\sigma)}\sum_{i=1}^3 u_{ii}^2
             \right]\diff\xx
 \label{eq_er}
\enq
The equation of motion for the displacement will then be,
\beq
 \rho \frac{\partial^2\uu}{\partial t^2} = -\frac{\delta}{\delta \uu(\xx)}(E_R[\uu] + E_I[\WW,\uu]).
 \label{eq_u_dyn}
\enq

It will prove useful to expand $\uu$ in terms of the elastic normal modes $\ffn$:
\beq
 \uu = \sum_{n} q^{(n)}(t)\ffn({\bf x})
 \label{eq_u_sum_f}
\enq
where the functions $\ffn$ satisfy the boundary conditions appropriate for $\uu$ and satisfy:
\begin{eqnarray}
  && \frac{\delta E_R[\ffn]}{\delta \ffn(\xx)} = 
            \omega^2_n\rho\ffn(\xx);\;\;\; n\in \Nint 
            \phantom{\frac{X}{X}}         \label{eq_modes_def1} \\
  && \frac{1}{M}\intvr\rho\ffn(\xx)\cdot{\bf f}^{(m)}(\xx)\diff\xx = \delta_{nm}   
            \phantom{\frac{X}{X}}         \label{eq_modes_def3}          
\end{eqnarray}
where $M$ is the total mass,
$
 M \df \intvr\rho\diff\xx.
$

In terms of the degrees of freedom, ${\qin}$, we have:
\beq
 \lag_I = -E_I = -B_1\sum_{n}\qin\sum_{i,j}\intvm\Omega_i\Omega_j\fnij\diff\xx
 \label{eq_ei_qin}
\enq
with
$$
 \fnij \df \frac{1}{2}\left(\frac{\partial\fin}{\partial x_j} + \frac{\partial\fjn}{\partial x_i}\right).
$$
We then see that the interaction Lagrangian is linear in the coordinates ${\qin}$, 
with coupling constants:
\beq
 \cin[\WW] \df \sum_{i,j}\intvm\Omega_i\Omega_j\fnij\diff\xx.
 \label{eq_def_cn}
\enq
This property will allow us to integrate out the reservoir degrees
of freedom to obtain an equation for the dynamics of the magnetization
in term of $\WW$ alone.

%
Let's first discuss the dynamics of the reservoir
degrees of freedom $\qin$.
Using equations \ceq{eq_u_dyn}-\ceq{eq_modes_def3} we
find the dynamical equations:
\beq
 \ddot q^{(n)} = -\omega^2_n\qin - \frac{B_1}{M}\cin[\WW].
 \label{eq_q_dyn}
\enq
Integrating \ceq{eq_q_dyn} we find
\begin{align}
 \qin(t) = & \qin|_0\cos(\omega_n t) + \frac{\dot q^{(n)}|_0}{\omega_n}\sin(\omega_n t) \nonumber \\
           &-\frac{B_1}{M\omega_n}\int_0^t\sin(\omega_n(t-t'))\cin[\WW(t')]\diff t',
 \label{eq_q_dyn_2}
\end{align}
where $\qin|_0$ and $\dot q^{(n)}|_0$ are the initial values
of $\qin$ and $\dot q^{(n)}$ respectively. 
The coupling of the magnetization to the reservoir will cause
damping and frequency renormalization. In order to be able
to separate the two effects is useful to integrate the last term
on the right hand side of \ceq{eq_q_dyn_2} by parts obtaining:
\begin{align}
  \qin(t) = & \qin|_0\cos(\omega_n t) + \frac{\dot q^{(n)}|_0}{\omega_n}\sin(\omega_n t)  \nonumber \\
            &-\frac{B_1}{M\omega^2_n}\cin[\WW(t)] +  
              \frac{B_1}{M\omega^2_n}\cin[\WW(0)]\cos(\omega_n t)  \nonumber \\
	    &+\frac{B_1}{M\omega^2_n}\int_0^t\diff t'\left[\cos(\omega_n (t-t'))
              \phantom{\frac{X}{X}}\right. \nonumber \\
            &\left.\cross\int_{V_M}\left.\frac{\delta\cin}{\delta\WW}\right|_{\xx',t'}
                                  \cdot\left.\frac{\partial\WW}{\partial t'}\right|_{\xx'}\diff\xx'\right].
  \label{eq_q_dyn_3}
\end{align}
Using the expression of the interaction Lagrangian given by \ceq{eq_ei_qin}
and the definition of the coupling constants $\cin$ we have:
\beq
 \frac{\delta \lag_I}{\delta\WW} = -B_1\sum_{n}\qin\frac{\delta \cin}{\delta\WW}.
 \label{eq_dei_dw_1}
\enq
Combining equations \ceq{eq_dyn_ww_gen}, \ceq{eq_q_dyn_3} and \ceq{eq_dei_dw_1} for the dynamics
of the magnetization we find: 
\begin{align}
  \frac{\partial\WW}{\partial t} =  
 &\WW\cross\frac{\gamma}{M_s}\frac{\delta E_S}{\delta\WW} +
  \WW\cross\frac{\gamma}{M_s}\frac{\delta\Delta\lag(\WW)}{\delta\WW}  \nonumber \\ 
 &+ \WW\cross\frac{\gamma}{M_s} \sum_{n}
    \left[ B_1 \left.\frac{\delta \cin}{\delta\WW}\right|_{\xx,t}\right. \nonumber \\
 &\times\left( \qin|_0\cos(\omega_n t) + \frac{\dot q^{(n)}|_0}{\omega_n}\sin(\omega_n t)
             \right) \nonumber \\
 &         - \frac{B_1^2}{M\omega^2_n}\cin[\WW(t)]
             \left.\frac{\delta \cin}{\delta\WW}\right|_{\xx,t}  \nonumber \\
 & +  \frac{B_1^2}{M\omega^2_n}\cin[\WW(0)]\cos(\omega_n t)
             \left.\frac{\delta \cin}{\delta\WW}
             \right|_{\xx,t} 
   \nonumber \\
  &+\frac{B_1^2}{M\omega^2_n}\int_0^t\diff t'\int_{V_M}\diff\xx'
   \cos(\omega_n (t-t'))
   \left.\frac{\delta\cin}{\delta\WW}\right|_{\xx',t'} \nonumber \\
  &\cdot
   \left.
   \left.\frac{\partial\WW}{\partial t'}\right|_{\xx'}
   \left.\frac{\delta \cin}{\delta\WW}\right|_{\xx,t}
   \right].   
 \label{eq_dyn_ww_3}
\end{align}
The counter term $\Delta\lag$ of the total Lagrangian is defined to cancel the 
frequency renormalizing term:
\beq
 \WW\cross \frac{\gamma}{M_s}\sum_{i,n}\frac{B_1^2}{M\omega^2_n}\cin[\WW(t)]
 \left.\frac{\delta \cin}{\delta\WW}\right|_{\xx,t}
 \label{eq_fr_ren}
\enq
It follows from Eq. \ceq{eq_def_cn} that 
\beq
 \frac{\delta \cin}{\delta\Omega_l} = \sum_i\Omega_i
                                      \left[\frac{\partial f_l^{(n)}}{\partial x_i} + \frac{\partial f_i^{(n)}}{\partial x_l}
				      \right]
\enq

To simplify and extract the physical content from these cumbersome equations, we identify
the memory friction kernel tensor $\gamma_{jm}$:
\begin{align}
 \gamma_{jm}
 (t,t',\xx,\xx') 
                     \df
                      \Theta(t-t') 
                      \sum_{n}&\frac{\gamma}{M_s} 
                      \frac{B_1^2}{M\omega^2_n}\cos(\omega_n (t-t')) \nonumber \\
                      &\times
                      \left.\frac{\delta\cin}{\delta\Omega_m}\right|_{\xx',t'}
                      \left.\frac{\delta \cin}{\delta\Omega_j}\right|_{\xx,t}  
 \label{eq_def_gamma_jl}
\end{align}
where $\Theta(t-t')$ is the Heaviside function.
We also recognize the random field ${\bf h}$:
\beq
 {\bf h}(\xx,t) \df  \frac{B_1}{M_s}\sum_{n}
                     \left[\qin|_0\cos(\omega_n t) + 
		       \frac{\dot q^{(n)}|_0}{\omega_n}\sin(\omega_n t)\right]
                     \frac{\delta \cin}{\delta\WW}.
 \label{eq_def_h}
\enq
Assuming that the distribution of initial positions of the environment degrees of 
freedom follows the canonical classical equilibrium
density for the unperturbed reservoir
we find that 
\begin{align}
 \langle\hh(\xx,t)\rangle &= 0  \label{eq_h_av_0},\\
 \langle h_j(\xx,t)h_m(\xx',t')\rangle &= \frac{2 K_B T}{\gamma M_s}\gamma_{jm}(t,t',\xx,\xx')
 \label{eq_h_av}.
\end{align}
In terms of $\gamma_{jm}$ and $\hh$ the dynamical equation for $\WW$
takes the form:
\begin{align}
  \frac{\partial\Omega_l}{\partial t} = 
 &\epsilon_{ijl}\Omega_i\frac{\gamma}{M_s}
  \frac{\delta E_S}{\delta\Omega_j} +
  \gamma\epsilon_{ijl}\Omega_i h_j  \nonumber \\
 &+\epsilon_{ijl}\Omega_i\int_0^t\diff t'\int_{V_M}\diff\xx'
  \sum_m\gamma_{jm}(t,t',\xx,\xx')\left.\frac{\partial\Omega_m}{\partial t'}\right|_{\xx'} 
  \nonumber \\
 &+\epsilon_{ijl}\Omega_i\frac{\gamma}{M_s}
  \sum_{i,n}
  \frac{B_1^2}{M\omega^2_n}\cin[\WW(0)]\cos(\omega_n t)\frac{\delta \cin}{\delta\Omega_j}.
  \nonumber
\end{align}
The final term is an artifact of the assumption 
that in the initial state the reservoir was decoupled
from the system \cite{weiss_book,hanggi}.  
Dropping this term, 
the dynamical equations for magnetization coupled to
a thermal bath of elastic modes is:
\begin{align}
  \frac{\partial\Omega_l}{\partial t} = 
  &\epsilon_{ijl}\Omega_i\frac{\gamma}{M_s}\frac{\delta E_S}{\delta\Omega_j} +
  \gamma\epsilon_{ijl}\Omega_i h_j \nonumber \\
  &+\epsilon_{ijl}\Omega_i\int_0^t\diff t'\int_{V_M}\diff\xx'
  \sum_m\gamma_{jm}(t,t',\xx,\xx')\left.\frac{\partial\Omega_m}{\partial t'}\right|_{\xx'}
  \label{eq_dyn_ww_4}
\end{align}
with $\gamma_{jm}$ defined by \ceq{eq_def_gamma_jl} and $\hh$ a random field
with statistical properties given by \ceq{eq_h_av_0} and \ceq{eq_h_av}.
Equation \ceq{eq_dyn_ww_4} is quite general.
In particular notice that to obtain \ceq{eq_dyn_ww_4} we didn't
perform any expansion in $\WW$. As a consequence, as long
as we keep the exact form for $E_S(\WW)$, equations 
\ceq{eq_dyn_ww_4} includes also the effects of spin wave
interactions.  In principle we could also include in $E_S$
a term to take into account the scattering of spin waves due to disorder. 
Equation \ceq{eq_dyn_ww_4} does not, however, 
take into account the coupling between the magnetization
and particle-hole excitations.  As we discuss in Section VI, this coupling
appears to be of critical importance in many metallic ferromagnets.

Equation \ceq{eq_dyn_ww_4} is very different from the 
standard stochastic Landau-Lifshitz-Gilbert (s-LLG) equation,
Eq. \ceq{eq_llg}. 
Because the magnetoelastic energy, $E_I$, \ceq{eq_def_ei},
is nonlinear in the magnetization, in \ceq{eq_dyn_ww_4} both
the damping kernel and the random field depend on the 
magnetization and therefore are state dependent.
This is in contrast with the s-LLG equation for 
which both the damping kernel, $\alpha\delta(t-t')$,
and the random field are independent of $\WW$.

Another difference between Eq. \ceq{eq_dyn_ww_4} and 
the s-LLG equation is that the damping kernel,
$\gamma_{jm}$, is in general a tensor.
The tensor character of the damping has been suggested 
previously on phenomenological grounds \cite{smith_jap_01}.
Starting from the physical coupling \ceq{eq_def_ei},
in our approach the tensor character of $\gamma_{jm}$
appears naturally as a consequence of: 
{\it(a)} the nonlinearity in  $\WW$ of the 
magnetoelastic coupling \ceq{eq_def_ei},
{\it (b)} the anisotropy of the
elastic modes due to the boundary conditions and/or
anisotropy of the elastic properties.
For small oscillations
of $\WW$ around its equilibrium
(up to quadratic order), the kernel $\gamma_{jm}$ can be assumed
to be independent of $\WW$. Even in this linearized case, the damping
kernel that appears in \ceq{eq_dyn_ww_4} will still
have a tensor form due to the anisotropy of the
elastic modes.

As mentioned above, the standard s-LLG damping kernel
is simply $\alpha\delta(t-t')$, {\em i.e.} the damping is
frequency-independent.  As a consequence, from the Fluctuation
Dissipation Theorem, we have that the spectrum of the
random field that appears in \ceq{eq_llg} is also frequency-independent.
This differs from equation \ceq{eq_dyn_ww_4} for which
the damping kernel, and therefore the spectrum of the random field,
is frequency-dependent.

Given the geometry and the material properties
of the system we can find the elastic modes,
$\ffn$, and then integrate
equation \ceq{eq_dyn_ww_4} using a micromagnetic approach.
The integration of equation \ceq{eq_dyn_ww_4} could 
give insight in particular on the damping of the
uniform magnetization mode for different geometries and 
show the range of validity of the classic picture
\cite{sparks_book}
of a two stage damping process in which the motion of the
coherent magnetization induces 
non uniform magnetic modes on short time scales that then decay to lattice vibrations.

\section{Magnetization coupled to elastic modes: uniform magnetization}
We now study the dynamics of the uniform magnetic mode
in the case when we can neglect its interaction with spin
waves and the only coupling to external degrees of freedom
is magnetoelastic.  Projecting Eq. \ceq{eq_dyn_ww_4} on the uniform mode we find that 
\begin{align}
 \frac{\diff\Omega_l}{\diff t} = &
 \epsilon_{ijl}\Omega_i\frac{\gamma}{V_M M_s}\intvm\frac{\delta E_S}{\delta\Omega_j}\diff\xx +
 \epsilon_{ijl}\Omega_i\frac{\gamma}{V_M}\intvm h_j\diff\xx  \nonumber \\
&+\epsilon_{ijl}\Omega_i\frac{1}{V_M}\int_0^t\hspace{-0.3cm}\diff t'
  \intvm\hspace{-0.4cm}\diff\xx\intvm\hspace{-0.4cm}\diff\xx'
  \sum_m\gamma_{jm}(t,t',\xx,\xx')\frac{\diff\Omega_m}{\diff t'}.
  \label{eq_dyn_ww_un}
\end{align}
Let's define the space averaged error field
$$
 \bar\hh(t)\df\frac{1}{V_M}\intvm \hh(\xx,t)\diff\xx,
$$
the damping kernel
$$
 \bar\gamma_{jm}(t,t')\df\frac{1}{V_M}\intvm\diff\xx\intvm\diff\xx'\gamma_{jm}(t,t',\xx,\xx'),
$$
and the coefficients
$$
c_{l}^{(n)}\df \intvm\frac{\delta\cin}{\delta\Omega_l}\diff\xx.
$$
Using the fact that $\WW$ is uniform we obtain
\beq
 c_{l}^{(n)} = \sum_i\Omega_i\intvm\left[\frac{\partial f_l^{(n)}}{\partial x_i} + \frac{\partial f_i^{(n)}}{\partial x_l}\right]\diff\xx.
 \label{eq_ciln}
\enq
In terms of the coefficients $c_{l}^{(n)}$ we can then write:
$$
 \bar h_l = \frac{B_1}{M_s V_M}\sum_{n} c_{l}^{(n)}
            \left(\qin|_0\cos(\omega_n t) + 
		  \frac{\dot q^{(n)}|_0}{\omega_n}\sin(\omega_n t)
            \right)
$$
and
\beq
 \bar\gamma_{jm} = \Theta(t-t')\frac{\gamma B_1^2}{M_s M V_M}
                   \sum_{n}\frac{1}{\omega_n^2}c_{j}^{(n)}(t) c_{m}^{(n)}(t')
		   \cos(\omega_n(t-t')).
 \label{eq_gamma_un}
\enq
The uniform magnetization dynamics can then be expressed in terms of the spatially
averaged random field $\bar\hh$ and memory friction kernel
$\bar\gamma_{jl}$:
\begin{align}
  \frac{\diff\Omega_l}{\diff t} = 
  &\epsilon_{ijl}\Omega_i\frac{1}{V_M}\frac{\gamma}{M_s} \intvm\frac{\delta E_S}{\delta\Omega_j}\diff\xx +
  \gamma\epsilon_{ijl}\Omega_i \bar h_j \nonumber \\
  &+\epsilon_{ijl}\Omega_i\int_0^t\diff t'\sum_m\bar\gamma_{jm}(t,t')\frac{\diff\Omega_m}{\diff t'}
  \label{eq_dyn_ww_un_2}
\end{align}
with
\begin{align}
 \langle\bar\hh\rangle &= 0 \label{eq_av_h} 
\end{align}
and
\begin{align}
 \langle \bar h_j(t)\bar h_m(t')\rangle &= \frac{2K_B T}{\gamma V_M M_s}
   \bar\gamma_{jm}(t,t'). \label{eq_corr_h}
\end{align}

\section{Thin Film Uniform Magnetization Dynamics}
\label{section_thin_film}
\begin{figure}[hbt]
  \begin{center}
  \includegraphics[height = 5cm]{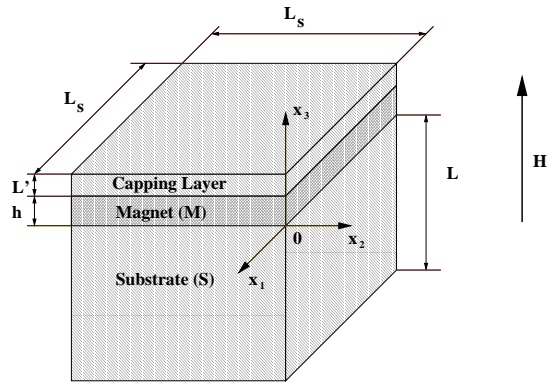}
  \end{center}
  \caption{Geometry considered for the case of a 
           thin ferromagnetic film on a non-magnetic substrate.}
  \label{fig_slabs} 
\end{figure}

We now apply equation \ceq{eq_dyn_ww_un_2} to study the dynamics
of the uniform magnetization in a thin ferromagnetic film
placed on top of a non magnetic substrate
and covered by a non magnetic 
capping layer, as illustrated in Fig. \ref{fig_slabs}.
We assume that all media are 
polycrystalline and treat them as isotropic.
We will assume the lateral size, $L_s$, Fig. \ref{fig_slabs},  
to be much bigger than the film 
thickness $h$.
Notice that if we take $L_s$ bigger than the domain wall width
our assumption that the non uniform magnetic modes are quenched
wouldn't be valid anymore.
We will consider only oscillations of the magnetization
around an equilibrium position parallel to the $x_3$ axis
so that we can calculate the damping kernel tensor $\gamma_{jm}$
assuming the elastic modes to depend only on $x_3$.
Otherwise, to find the correct damping kernel,
we would have to take into account the fact that the
lateral size, $L_s$, is finite and solve the full 3D elasticity problem 
for the elastic modes.
%
%
\subsection{Damping kernel and random field}
To find the dynamics of the magnetization using
equation \ceq{eq_dyn_ww_un_2} we need to evaluate the
memory friction kernel $\gamma_{jm}$.  The first step in
this calculation is the determination of the elastic normal modes $\ffn$
which satisfy the following equation:
\beq
  \omega_n^2\rho\ffn = -\frac{E}{2(1+\sigma)}\nabla^2\ffn -
                   \frac{E}{2(1+\sigma)(1-2\sigma)}\nabla(\nabla\cdot\ffn).
 \label{eq_pde_ffn}
\enq
We allow the film, the
substrate, and the capping layer to have different elastic properties and  
solve equation \ceq{eq_pde_ffn} separately in the different subsystems using 
the appropriate elastic constants.
We assume for the sake of definiteness that the substrate and
capping layer material is identical.  
We then match solutions by imposing the continuity of displacement and stresses
at the interfaces $x_3 = 0$, and $x_3 = h$.
As boundary conditions we assume the top surface of the capping
layer to be free and no displacement at the bottom
of the substrate.

Because in our case the elastic modes only depend on $x_3$,
Eq. \ceq{eq_ciln} simplifies to 
$$
 c_{l}^{(n)} = L_s^2\sum_i\Delta\fin[\delta_{il}\Omega_3 +\Omega_i\delta_{3l}]
$$
with
$$
 \Delta\fin \df \fin(h) - \fin(0).
$$
The spatially averaged damping coefficients have a simple expression in
terms of the $\Delta\fin$: 
\begin{align}
 \bar\gamma_{jl} =& \Theta(t-t')\frac{L_s^2B_1^2}{M h}
                   \sum_{n}\frac{[\Delta\fin]^2}{\omega_{n}^2}\cos(\omega_{n}(t-t')) \nonumber \\
		  &\times [\delta_{ij}\Omega_3(t)  +\Omega_i(t) \delta_{3j}]
		   [\delta_{il}\Omega_3(t') +\Omega_i(t')\delta_{3l}].
 \label{eq_gamma_un_film}
\end{align}
Eq. \ref{eq_gamma_un_film} follows from the completeness relation of the polarization vectors.
Once we know the coefficients $\Delta\fin$, Eqs.
\ceq{eq_gamma_un_film},
\ceq{eq_av_h}, and 
\ceq{eq_corr_h} 
completely specify the dynamical equation \ceq{eq_dyn_ww_un_2} for the
magnetization.

As an example we consider the case of a polycrystalline
ferromagnetic thin film, like YIG, placed on a substrate of 
a polycrystalline paramagnet like Tantalum, {\it Ta}.
As typical values we take \cite{bozorth_book} the ones
listed in Table \ref{table1}. 
\begin{table}[!!!!!!!!!!!!!!!!!h]
 \caption{Elastic properties. $c_t$, $c_l$ are the transverse and longitudinal
          speed of sound respectively.}
 \centering
 \begin{tabular}{|c|c|c|}
  \hline
  \phantom{aaaa} & \phantom a Magnetic Film \phantom a & Substrate/Capping Layer \\
  \hline\hline
  $E$         &   200 Gpa      &    180 Gpa              \\
  $\sigma$    &   0.33         &    0.33                 \\
  $\rho$      &   5.0 g/cm$^3$ &    16.6 g/cm$^3$        \\
  $c_t$       &   4.0 km/s     &    2.0 km/s             \\
  $c_l$       &   5.0 km/s     &    4.1 km/s             \\
  \hline
 \end{tabular}
 \label{table1}
\end{table}
For the magnetostriction we assume $\lambda= 2\times 10^{-6}$.
Using equation \ceq{eq_b1}, we find that $B_1 = 4\times 10^6 \text{ergs/cm}^3$.
Given the elastic modes implied by these parameter values, we can  
calculate the coefficients $\Delta\fin$. 
Once we know the coefficients $\Delta\fin$ we have all the elements
to completely specify equation \ceq{eq_dyn_ww_un_2}.

%
%
%
We generate a  
stochastic field $\bar\hh$ with the correct statistical properties 
by using its Fourier representation.  To obtain 
\beq
 \langle y(t)y(t')\rangle = G(t-t')
 \label{eq_corr_y}
\enq
we choose \cite{gardiner_book}
\beq
 \langle y(\omega)y(\omega')\rangle = \delta(\omega-\omega')G(\omega)
 \label{eq_corr_yw}
\enq
where
$$
 y(\omega) = \frac{1}{2\pi} \int_{-\infty}^{\infty}y(t)\ee^{-i\omega t}\diff t
$$
and 
$$
 G(\omega) = \frac{1}{2\pi} \int_{-\infty}^{\infty}G(\tau)\ee^{-i\omega \tau}\diff \tau.
$$

\begin{figure}[hbt]
 \begin{center}
 \subfigure[]
 { 
  \includegraphics[width = 8.0cm]{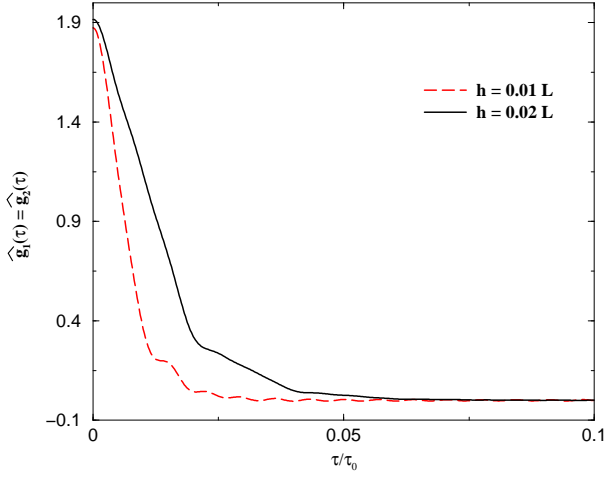}
  \label{fig_g12_t}
 }
 \subfigure[]
 { 
  \includegraphics[width = 8.0cm]{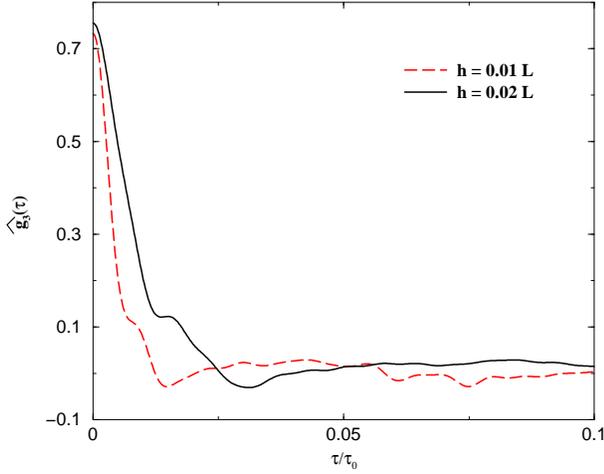}
  \label{fig_g3_t}
 } 
 \end{center}
 \caption{(Color online) Profiles of $\hat g_1 \df g_1(\tau)c^2/[h(L+h+L')]$, (a), 
          and $\hat g_3 \df g_3(\tau)c^2/[h(L+h+L')]$, (b), for the 
          case of a thin magnetic film on a Tantalum substrate;
	  $\tau_0\df L/c_{t,M}$.
	  For the standard s-LLG equation $g_i(\tau)$ would simply be
	  a Dirac's delta centered at $\tau = 0$.
         }
\end{figure}

\begin{figure}[hbt]
 \begin{center}
 \subfigure[]
 { 
  \includegraphics[width = 8.0cm]{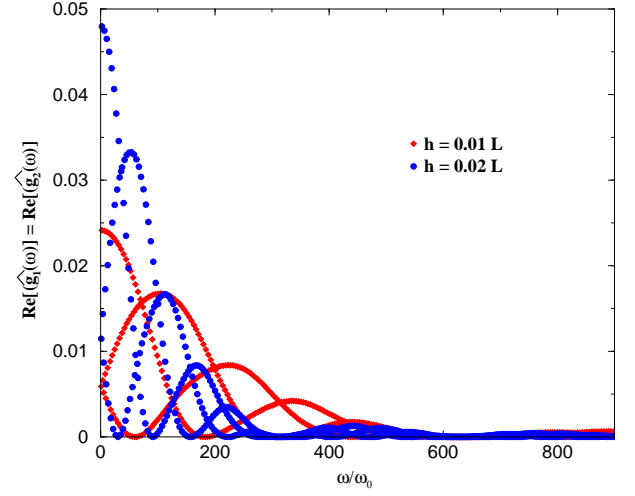}
  \label{fig_re_g12_w}
 }
 \subfigure[]
 { 
  \includegraphics[width = 8.0cm]{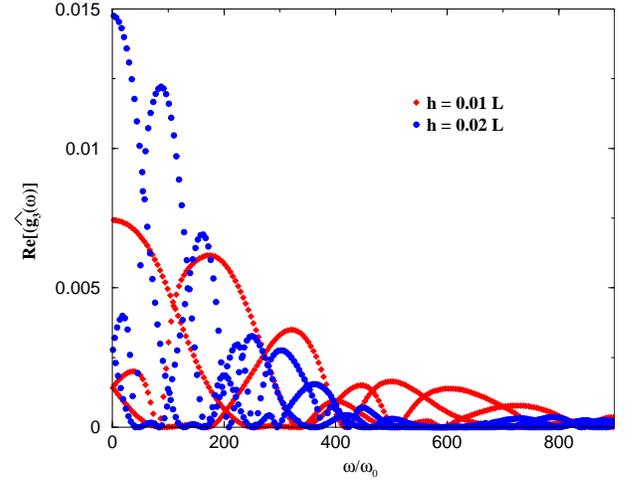}
  \label{fig_re_g3_w}
 }
 \end{center} 
 \caption{(Color online) 
          Values of ${\rm Re}[\hat g_1(\omega)]\df{\rm Re}[g_1(\omega)] c^2/[h(L+h+L')]$ (a)
          and ${\rm Re}[\hat g_3(\omega)]\df{\rm Re}[g_3(\omega)] c^2/[h(L+h+L')]$ (b) 
          at the elastic modes frequencies $\{\omega_n\}$ for the case of
          a thin magnetic film on a Tantalum substrate.
	  Shown are the values for $h=0.01L$, diamonds, and $h=0.02L$, circles.
          For any $\omega_n$ ${\rm Re}[\hat g_i(\omega_n)]$ is unique
	  even though this is not completely evident from the figure because
	  in order to show the behavior of the auxiliary kernels 
          over a wide frequency range, the resolution is not high enough
          to always show the separation between the single points. 
	  For the standard s-LLG equation $g_i(\omega)$ would simply be
	  a constant.
         }
\end{figure}

In our case we have from Eq. \ceq{eq_gamma_un}, that the memory friction
kernel $\bar\gamma_{jl}$ depends separately on $t$ and $t'$. 
As a consequence, through \ceq{eq_corr_h}, we have that 
the average $\langle\bar\hh(t)\bar\hh(t')\rangle$ doesn't depend only
on the time difference $\tau=t-t'$. The random field 
$\bar\hh(t)$ therefore doesn't define an ergodic process
and in particular we cannot use equation \ceq{eq_corr_yw}.
For this reason it is convenient to define the auxiliary
random variables:
$$
 x_i \df \sum_n \Delta\fin
         \left[\qin|_0\cos(\omega_n t) + 
		  \frac{\dot q^{(n)}|_0}{\omega_n}\sin(\omega_n t)
            \right]
$$
and the auxiliary kernels:
$$
 g_i(t-t')\df \Theta(t-t')\sum_n\frac{[\Delta\fin]^2}{\omega_n^2}\cos(\omega_n(t-t'))
$$
so that we have:
$$
 \langle x_i(t)x_j(t')\rangle = \frac{2 K_B T}{M}g_i(t-t')\delta_{ij}.
$$
The random variables $x_i(t)$ therefore describe an ergodic process
and we can use equation \ceq{eq_corr_yw} to generate them.
In terms of $x_i$ and $g_i$ we have:
$$
 \bar h_l = \frac{B_1}{M_s h}\sum_{i}x_i[\delta_{il}\Omega_3(t') +\Omega_i(t')\delta_{3l}];
$$
\begin{align}
 \bar\gamma_{jm} = &\frac{\gamma L_s^2B_1^2}{M_s M h}
                   \sum_{i}g_i
		   [\delta_{ij}\Omega_3(t)  +\Omega_i(t) \delta_{3j}] \nonumber \\
		   &\times[\delta_{im}\Omega_3(t') +\Omega_i(t')\delta_{3m}].
 \label{eq_gamma_thin_film}
\end{align}
To generate the random field 
and calculate $\bar\gamma_{jl}$ 
we then
have to calculate the quantities $g_i(\tau)$ and their Fourier transforms
$g_i(\omega)$. 
Figures \ref{fig_g12_t}-\ref{fig_re_g3_w}
show some typical profiles
for $g_i(\tau)$ and $g_i(\omega)$ using for the mechanical
properties the values of table \ref{table1}.
We find that in general
$g_i(\tau)$ doesn't depend on the thickness of the capping layer
$L'$.

In the limit in which we can linearize the magnetoelastic interaction with
respect to $\WW$, we have:
\beq
 \bar\gamma_{jm}(\tau) = \frac{\gamma B_1^2 L_s^2}{M_s M h}g_j(\tau)\delta_{jm}.
 \label{eq_gamma_lin_ei}
\enq
The damping kernel is diagonal with
components equal, apart from an overall constant,
to $g_j(\tau)$, in contrast to the s-LLG equation for
which we have $\bar\gamma_{jl}(\tau) = \alpha\delta(\tau)\delta_{jl}$.
The power spectrum of the random field component, $h_j$, 
is then proportional to $g_j(\omega)$,  in contrast to
the s-LLG equation for which the power spectrum of each
component $h_j$ is simply a constant.
Notice that even in this limit $\bar\gamma_{jm}$ preserves its
tensor form due to the anisotropy of the elastic modes.
In our specific case we have
$g_1=g_2\neq g_3$ due to the difference between 
the transverse and longitudinal speeds of sound.

From figures \ref{fig_g12_t}, \ref{fig_g3_t}, we see that 
$g_i(\tau)$ goes to zero for times longer than
$\tau_D \approx 5\;10^{-2}\tau_0=5h/c$, 
where $c\df c_{t,M}$ is the transverse speed of
sound in the magnet.
For a film 20 nm thick we then find $\tau_D\approx 10{\rm ps}$ .
When the relevant frequencies of $\WW$ are much lower
than $1/\tau_D$, we can replace the damping kernel
given by \ceq{eq_gamma_lin_ei} with the simple
kernel
$$
 \gamma_{jm} = \gamma_{j\rm eff}\delta(\tau)\delta_{jm}
$$
with $\gamma_{j\rm eff}$ given by:
\beq
 \gamma_{j\rm eff} = \frac{\gamma B_1^2 L_s^2}{M_s M h}\int_0^{\infty}g_j(\tau)\diff\tau.
 \label{eq_gamma_eff}
\enq
In this limit we recover a damping kernel of the same
form as the one that appears in the s-LLG equation.
Here $\gamma_{j\rm eff} $ is the equivalent to $\alpha$
in \ceq{eq_llg}.
From the results shown in Fig.\ref{fig_g12_t}, \ref{fig_g3_t}
we see that we have
$$
 \int_0^{\infty}g_j(\tau)\diff\tau\approx
 \frac{h^2 (L+h+L')}{c^3}
$$
and then:
\beq
 \gamma_{j\rm eff} = \frac{\gamma B_1^2 h}{M_s\rho c^3}. 
 \label{eq_gamma_estimate}
\enq
We find that the damping of magnetic modes in thin
films is proportional to $B_1^2 h$.
Assuming the values given in Table \ref{table2}
\begin{table}[!!!!!!!!!!!!!!!!!h]
 \caption{}
 \centering
 \begin{tabular}{|c|c|}
  \hline
  Quantity & Value \\
  \hline\hline
  $\gamma$         &   $1.76\times 10^7 {\rm s}^{-1}{\rm G}^{-1}$     \\
  $B_1$            &   $4\times 10^6 {\rm ergs/cm}^3;$ \\
  $M_s$            &   $150 {\rm G}$ \\
  $L$              &   $1 \mu{\rm m}$ \\
  $h$              &   $20 {\rm nm}$  \\
  \hline
 \end{tabular}
 \label{table2}
\end{table}
we find $\gamma_{1 \rm eff} =\gamma_{2 \rm eff} \approx 2\times 10^{-4}$. 
\clearpage

%
\subsection{Integration}
After generating the random field $\bar\hh$ in the way described
above we can proceed in integrating equation \ceq{eq_dyn_ww_un_2}.
We assume 
$\delta E_S/\delta\WW = -V_M M_s\HHeff$ with 
$\HHeff=(0,0,\Heff)$ and $\Heff$ simply a constant.
Let's define the dimensionless quantities:
$$
 \hat t  \df \gamma\Heff t;                                                \hspace{0.5cm}
 \hat{\rm \bf H}_{\rm eff} \df \frac{\HHeff}{\Heff};    \hspace{0.5cm}
 \hat\hh \df \frac{\hh}{\Heff}; \hspace{0.5cm}
$$
$$
 \hat\gamma_{jm} \df \frac{\bar\gamma_{jm}}{\gamma\Heff}; \hspace{0.5cm}
 \hat T \df \frac{2 K_B T}{\Heff M_s V_M};
$$
then in dimensionless form equation \ceq{eq_dyn_ww_un_2} takes the form,
\begin{align}
  \frac{\diff\Omega_l}{\diff \hat t} = 
 &-\epsilon_{ijl}\Omega_i \hat H_{{\rm eff} j} +
  \epsilon_{ijl}\Omega_i \hat h_j \nonumber \\
 &+ \epsilon_{ijl}\Omega_i\int_0^{\hat t}\diff \hat t'
                        \sum_m\hat\gamma_{jm}(\hat t,\hat t')\frac{\diff\Omega_m}{\diff \hat t'}
  \label{eq_dyn_ww_un_dimless}
\end{align}
with
\beq
 \langle \hat h_j(\hat t)\rangle = 0; \hspace{0.5cm}
 \langle \hat h_j(\hat t) \hat h_m(\hat t')\rangle =
 \hat T \hat\gamma_{jm}(\hat t, \hat t').
 \label{eq_stat_h_ww_dless}
\enq

Similarly, for $\delta E_S/\delta\WW = -V_M M_s\HHeff$, the standard
s-LLG equation, \ceq{eq_llg}, for the uniform mode, takes the dimensionless
form:
\beq
 \frac{\diff \WW}{\diff \hat t} = 
 -\WW\cross\hat{\rm \bf H}_{\rm eff} + \WW\cross\hat\hh  
 +\alpha\WW\cross\frac{\diff\WW}{\diff \hat t}
 \label{eq_llg_dimless}
\enq
with
\beq
 \langle \hat h_j(\hat t)\rangle = 0; \hspace{0.5cm}
 \langle \hat h_j(\hat t) \hat h_m(\hat t')\rangle =
 \alpha\hat T \delta(\hat t- \hat t').
 \label{eq_stat_h_llg_dless}
\enq

\begin{figure}[hbt]
  \begin{center}
  \subfigure[]
  {
   \includegraphics[width = 8.5cm]{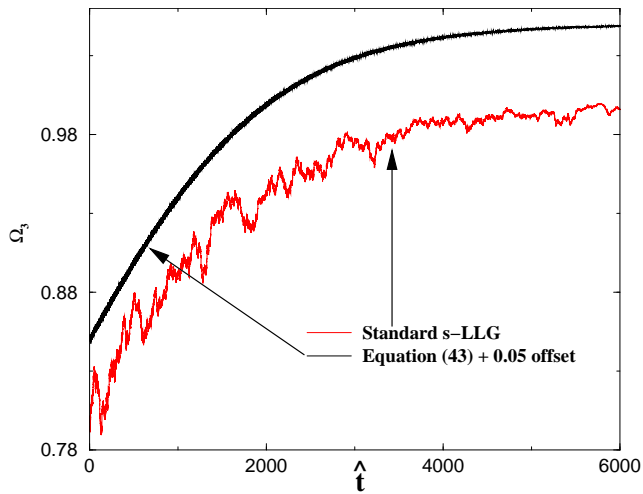}
   \label{fig_int_results1}
  }
  \subfigure[]
  {
   \includegraphics[width = 8.5cm]{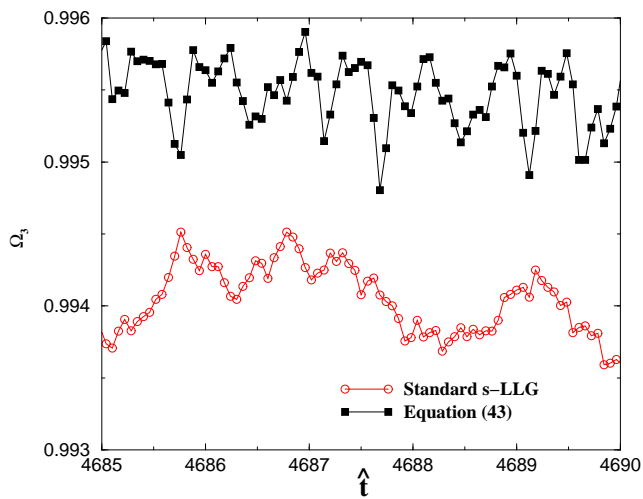}
   \label{fig_zoom_Om_3}
  }
  \caption{(Color online) $\Omega_3$ as a function of time obtained integrating the
  standard s-LLG equation, \ceq{eq_llg_dimless}, and equation \ceq{eq_dyn_ww_un_dimless}.
  In (a) the trace obtained using equation \ceq{eq_dyn_ww_un_dimless} has been offset
  up by $+0.05$ for clarity. In (b) the trace of $\Omega_3$ is shown on a short
  time scale.}
  \end{center}
\end{figure}

\begin{figure}[hbt]
  \begin{center}
  \subfigure[]
 { 
  \includegraphics[width = 8.5cm]{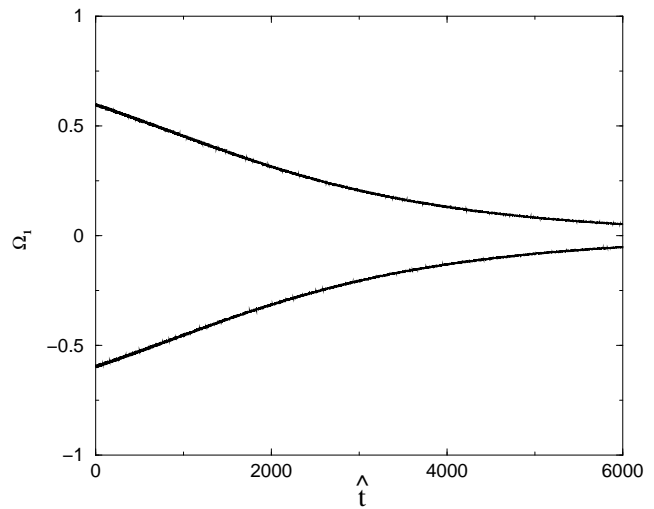}
  \label{fig_int_results2} 
 }
 \subfigure[]
 { 
  \includegraphics[width = 8.5cm]{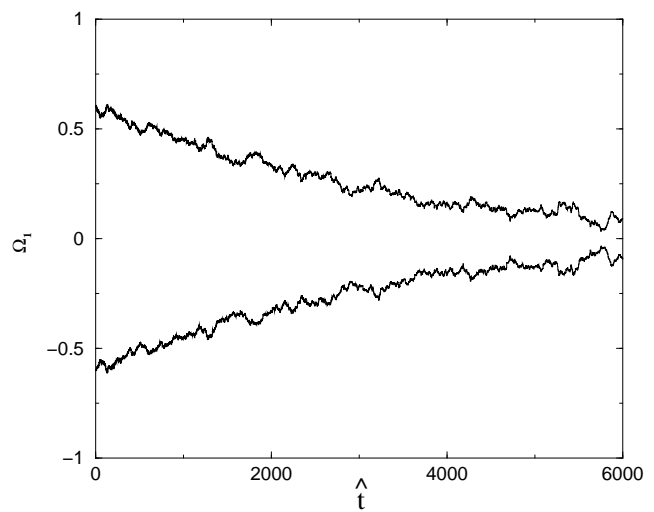}
  \label{fig_int_results3}
}
\caption{Envelope curves of the trace of $\Omega_1$ in time
         as obtained integrating equation \ceq{eq_dyn_ww_un_dimless}
	 (a) and equation \ceq{eq_llg_dimless} (b).
         $\Omega_1$ oscillates between the maximum and minimum value
         given by the envelope curves with frequency $\gamma H_{\rm eff}$,
	 equal to 1 in the dimensionless units used. 
	 }
  \end{center}
\end{figure}

Using for $\bar\gamma_{jm}$ the expression \ceq{eq_gamma_thin_film}
and for $g_i(\tau)$, $g_i(\omega)$ the results shown in figures
\ref{fig_g12_t}-\ref{fig_re_g3_w} and assuming $\hat T = 10^{-2}$
and the values given in Table \ref{table2}
we integrate equation  \ceq{eq_dyn_ww_un_dimless}.
We used the {stochastic Heun scheme}
that ensures convergence to the Stratonovich solution even in the
limit of zero autocorrelation time for the random field \cite{palacios_prb_98}.
The results of the integration 
are shown in figures \ref{fig_int_results1},\ref{fig_zoom_Om_3},\ref{fig_int_results2}.
As initial condition we took $\WW = (0.6,0,0.8)$, $\diff\WW/\diff\hat t = 0$.

We then integrated equation \ceq{eq_llg_dimless}
setting $\alpha = \gamma_{1 \rm eff}$ with $\gamma_{1 \rm eff}$
calculated using \ceq{eq_gamma_eff}. The results of the integration
are shown in figures \ref{fig_int_results1},\ref{fig_zoom_Om_3},\ref{fig_int_results3}.

From figures
\ref{fig_int_results1}-\ref{fig_int_results3} we see that
on average equation \ceq{eq_dyn_ww_un_dimless} and \ceq{eq_llg_dimless}
give very similar results. This is expected because for the
initial conditions chosen we are in the limit of
small oscillations around the equilibrium position and therefore
the dependence of $\hat\gamma_{jm}$ on $\WW$ is negligible.
The main differences, for the case considered, between the
results obtained using \ceq{eq_dyn_ww_un_dimless} and \ceq{eq_llg_dimless}
are in the random fluctuations of $\WW$.
This is a consequence of the different correlation in time
of the random field $\hh$ used in \ceq{eq_dyn_ww_un_dimless} and \ceq{eq_llg_dimless}.
For example we notice that
equation \ceq{eq_dyn_ww_un_dimless} seems to give
a less noisy dynamics than \ceq{eq_llg_dimless}
even though for both simulation $|\hat h|^2$ is of the same order of magnitude.
If we zoom on a short time interval, fig. \ref{fig_zoom_Om_3},
as a matter of fact, we see that on very short time scales the amplitude
of the random fluctuations for the two simulations is very
similar. However for \ceq{eq_llg_dimless} 
fluctuations with the same sign are much more likely
than for \ceq{eq_dyn_ww_un_dimless}.
This is due to the different spectral density of the random
field. For \ceq{eq_llg_dimless} we simply have 
$|\bar h_j(\omega)|^2 = \alpha\bar T$, whereas for
\ceq{eq_dyn_ww_un_dimless} $|\bar h_j(\omega)|^2$
is equal to $g_j(\omega)$  (considering that for our
simulation, to a good approximation, we can neglect the
dependence of the random field on $\WW$).
In particular  for \ceq{eq_dyn_ww_un_dimless} 
$|\bar h_j(\omega)|^2$ has a low frequency cutoff
at $\omega =\omega_0\df c_{t,M}/L$, where $c_{t,M}$
is the transverse speed of sound in the magnet. 
This implies that for \ceq{eq_dyn_ww_un_dimless} we have a much
lower probability than for \ceq{eq_llg_dimless}
to have consecutive fluctuations of the random
field with the same sign with the result that
the dynamics appears less noisy.
%

%
%
\section{Discussion and conclusions}
In this paper we derived the equation for the dynamics of the
magnetization taking into account its coupling 
to the lattice vibrations. The equation that we obtain,
\ceq{eq_dyn_ww_4}, is quite general. 
Equation \ceq{eq_dyn_ww_4} will have the same form
also if we include spin-spin and 
spin-disorder interactions.
To take into account these phenomena 
it is necessary only to add the appropriate terms
to the energy functional $E_S[\WW]$.

From the general equation we derived the equation,
\ceq{eq_dyn_ww_un_2}, for the dynamics of the uniform
magnetic mode in a thin magnetic film when nonuniform
magnetic modes can be assumed frozen out.
We find that in general the random field
that appears in the dynamical equation for the
magnetization has a correlation time, $\tau_D$,
of the order of the ratio between the film thickness, $h$,
and the sound velocity $c$.
When the timescale for the dynamics of the 
magnetization is much longer that $\tau_D$,
we recover the stochastic LLG equation.
In this limit we calculated the value of the 
effective Gilbert damping constant, $\alpha$.
For typical ferromagnetic insulators, like YIG,
we find $\alpha\approx 10^{-4}$, in good agreement
with the values measured in experiments \cite{sparks_book,Chen:1993}. 
We can then conclude
that for magnetic insulators magnetoelastic
coupling is the main source of magnetization
damping. 

Our work predicts that magnetic resonance experiments
on ferromagnetic insulators should be able to observe
the anisotropy of the damping and as a consequence
of the correlation of the thermal fluctuations. 
With our theory is possible to exactly
calculate the spectral density of the thermal fluctuations.
The spectral densities for small samples will
be different from the one observed in bulk experiments
because of the discreteness of the elastic modes.
It would be very interesting to test these results
with new experiments on small ferromagnetic insulators samples.
In particular for thin films one experimental consequence
of our work is that the correlation time of the 
magnetic fluctuations will be of the order
of $h/c$ where $h$ is the thickness of the ferromagnetic
film and $c$ the speed of sound in the magnet.
We also found that in the limit when the magnetization
evolves on time scales much bigger than $h/c$ 
the damping of the magnetic modes is directly
proportional to $B_1^2 h$. The line width of the
ferromagnetic resonance peak in
insulating ferromagnetic thin films should therefore scale as $B_1^2h$,
which, in principle, can be confirmed experimentally.

For ferromagnetic metals, like permalloy, we 
also find  $\alpha\approx 10^{-4}$. This value is about two orders
of magnitude smaller than the value observed experimentally
\cite{covington_prl_02}. The reason is that in
ferromagnetic metals the electronic degrees of freedom
are the main source of dissipation for the magnetization
\cite{Kittel:1956,Turov:1966}.
Starting from a model of localized d spins exchange-coupled
to s-band electron, the interaction Lagrangian will be:
$$
 \lag_I = J_{sd}\int\diff\xx\WW(\xx)\cdot{\bf s}(\xx)
$$
where $J_{sd}$ is the exchange coupling constant and
{\bf s} is the conduction electrons spin density:
$$
 {\bf s}(\xx) = \frac{1}{2}\sum_{a,b}
                \Psi^\dagger_a(\xx)
                {\boldsymbol \tau}_{ab}
                \Psi_{b}(\xx)
$$
where $\Psi$ are the s-band carrier field operators and 
${\boldsymbol \tau}_{ab}$ the representation
of the spin operator in terms of Pauli matrices.
By integrating out the s-band degrees of freedom,
in the linear response approximation Sinova et al.
\cite{Sinova:2004}, 
for the damping of the uniform magnetic mode find:
\begin{align}
 \alpha = &\lim_{\omega\to 0}
          \frac{g\mu_B J_{sd}^2}{2 M_s \hbar\omega}
	  \int\frac{\diff^3k}{(2\pi)^3}
	  \sum_{a,b}|\langle\psi_a(\kk)|\tau^+|\psi_b(\kk)\rangle|^2 \nonumber \\
	  &\times\int\frac{\diff\epsilon}{2\pi}A_{a,\kk}(\epsilon)A_{b,\kk}(\epsilon + \hbar\omega)
	  [f(\epsilon) - f(\epsilon + \hbar\omega)]
 \label{eq_alpha_metals}
\end{align}
where $A_{a,\kk}(\epsilon)$ and $A_{b,\kk}(\epsilon)$
are the spectral functions for s-band quasiparticles
and $f(\epsilon)$ is the Fermi-Dirac distribution.
Equation \ceq{eq_alpha_metals} gives zero damping
unless there is a finite-measure Fermi surface
area with spin degeneracy or there is a broadening
of the spectral function due to disorder
\cite{Tserkovnyak:2004:apl}.
Characterizing the quasiparticle broadening by
a simple number $\Gamma\df \hbar/\tau_s$, where
$\tau_s$ is the quasiparticle lifetime, we can assume:
\beq
 A_{a,\kk}(\epsilon)=\frac{\Gamma}{(\epsilon-\epsilon_{a,\kk})^2 + \Gamma^2/4}.
 \label{eq_Aa}
\enq
Inserting this expression for the spectral functions
in \ceq{eq_alpha_metals} we find $\alpha$
as a function of the phenomenological scattering 
rate $\Gamma$. 
Notice that \ceq{eq_alpha_metals} includes the contribution
both of intra-band, 
and  inter-band 
\cite{Heinrich:1967,Kambersky:1970,Berger:1977} 
quasiparticles scattering events. 
The intra-band contribution is due to spin-flip scattering
within a spin-split band and is nonzero only when intrinsic
spin-orbit coupling is present. 
From equation \ceq{eq_alpha_metals}, using the expression
for $A_{a,\kk}(\epsilon)$ given in \ceq{eq_Aa}, we see
that in the limit of weak disorder, small $\Gamma$,
the intra-band contribution to $\alpha$
is proportional to $1/\Gamma$, in agreement with
experimental results for clean ferromagnetic metals
with strong spin-orbit coupling
\cite{Bhagat:1966,Bhagat:1974,Heinrich:1979,Rudd:1985}
and previous theoretical work
\cite{Kambersky:1970,Korenman:1972,Korenman:1974,Kunes:2002,Kunes:2003,Berger:1977}.
Similarly  from \ceq{eq_alpha_metals} we  see that 
the inter-band contribution to $\alpha$
is proportional to $\Gamma$. This result 
agrees with the experimental results for ferromagnetic
metals with strong disorder \cite{Ingvarsson:2002} and
previous theoretical work 
\cite{Heinrich:1967,Kambersky:1970,Berger:1977}.
Notice that equation \ceq{eq_alpha_metals} implicitly
also includes the contribution due to the so called
spin-pumping effect
\cite{Berger:1996,Tserkovnyak:2002,Simanek:2003,Simanek:2003:b,Rebei:2004}
in which spins are transferred from the ferromagnetic
film to adjacent normal metal layers as a consequence
of the precession of the magnetization.
In order to calculate this effect in first approximation
we simply have to substitute in \ceq{eq_alpha_metals}
the conduction band quasiparticle states, $\psi$,  
calculated taking into account the heterogeneity
of the sample.
Assuming for the scattering rate, $1/\tau_s$, typical
values estimated by transport experiments, 
from equation \ceq{eq_alpha_metals} we find values
of $\alpha$ in good agreement with experiments.

In summary we have studied in detail the effect of the 
magnetoelastic coupling to the dynamics of the magnetization.
Starting from a realistic form for the magnetoelastic coupling
we have found the expression for the damping kernel,  $\gamma_{jm}$.
We find that in general $\gamma_{jm}$ 
is a nondiagonal tensor non-local in time and space.
The knowledge of the exact expression of $\gamma_{jm}$ 
allows us to correctly take into account 
the autocorrelation of the noise term overcoming
the zero correlation approximation of the 
stochastic Landau-Lifshitz-Gilbert equation.
We find that for thin films for which the single
domain approximation is valid, both the damping and 
the fluctuations correlation time are proportional
to the film thickness.
Our results apply to systems for which the direct 
coupling of the magnetization to the lattice vibrations
is the main source of the magnetization relaxation.
We have shown that this is the case for ferromagnetic
insulators  whereas for ferromagnetic metals 
the magnetization relaxation is mainly due to the s-d exchange coupling.

%
%
\section{Acknowledgments}
It is a pleasure to thank Harry Suhl, Thomas J. Silva, Alvaro S. N\'u\~nez 
and Joaqu\'{\i}n Fern\'andez-Rossier for helpful discussions. 
This work was supported by the Welch Foundation, by the 
National Science Foundation under grants 
DMR-0115947 and DMR-0210383, and by a grant from Seagate 
Corporation.
%
%
\appendix
\section{Simple Estimate of $\gamma_{\rm eff}$}

%
Let's start from the definition of $\bar\gamma_{jm}$ (equation \ceq{eq_gamma_un}):
\beq
 \bar\gamma_{jm} = \Theta(t-t')A_1
                   \sum_{n}\frac{1}{\omega_n^2}c_{j}^{(n)}(t) c_{m}^{(n)}(t')
		   \cos(\omega_n(t-t'))
\enq
where 
$
 A_1 \df \gamma B_1^2/M_s M V_M
$.
For the case of thin film we found:
$$
 c_l^{(n)} = L_s^2\sum_i\Delta f_i^{(n)}[\delta_{il}\Omega_3 + \Omega_i\delta_{3l}]
$$
where:
$$
 \Delta f_i^{(n)} \df f_i^{(n)}(h) - f_i^{(n)}(0).
$$
Notice that, by definition, $f_i^{(n)}$ are dimensionless and so are
the quantities $\Delta f_i^{(n)}$.
Assuming that at equilibrium is $\WW=(0,0,1)$ and keeping only the leading terms in
$\WW$ in the expression for $c_l^{(n)}$ we have:
$$
 c_l^{(n)} = L_s^2\Delta f_l^{(n)}.
$$
Let's now expand the collective index $n$ in its components:
${\bf k}, s$ where $s$ is the polarization index of the elastic modes.
Then, using the completeness of the polarization vectors and the fact that
the polarization directions are parallel to the axis $x_1, x_2, x_3$ 
we have:
\begin{align}
 \bar\gamma_{jm} =& \Theta(t-t')A_1
                    \sum_{n}\frac{1}{\omega_n^2}c_{j}^{(n)}(t) c_{m}^{(n)}(t')
		    \cos(\omega_n(t-t'))  \nonumber \\
                 =& \Theta(t-t')A_1 L_s^4
		    \sum_{{\bf k},s}\frac{1}{\omega_{{\bf k},s}^2}\Delta f_j^{{\bf k},s} \Delta f_m^{{\bf k},s} 
		    \cos(\omega_{{\bf k},s}(t-t'))  \nonumber \\
		 =& \Theta(t-t')A_1 L_s^4
		    \sum_{{\bf k}}\frac{1}{\omega_{{\bf k},j}^2}\Delta f_j^{{\bf k}} \Delta f_m^{{\bf k}}\delta_{jm} 
		    \cos(\omega_{{\bf k},j}(t-t'))  \nonumber 
\end{align}
Now note that:
$$
 M   = \rho L_s^2 L (1 + \hat h + \hat L'); \hspace{1cm}
 V_M = L_s^2 h; 
$$
where $\hat h \df h/L $,$\hat L' \df L'/L$.
Then we can write:
\begin{align}
  \bar\gamma_{jm} = &\Theta(t-t')\frac{\gamma B_1^2}{M_s \rho L(1 + \hat h + \hat L') h}\nonumber \\
		    &\times\sum_{{\bf k}}\frac{[\Delta f_j^{{\bf k}}]^2 }{\omega_{{\bf k},j}^2}\delta_{jm} 
		    \cos(\omega_{{\bf k},j}(t-t')).  \nonumber 
\end{align}
For small enough $h/L$ we can assume
$
 \Delta f_j^{(k)} \approx kh
$
with a cutoff for $k_D$ such that $k_D h = 1$.We can then define
the cutoff frequency $\omega_D \df c k_D=c/h$.
With this approximation we have:
\begin{align}
 \sum_{{\bf k}}\frac{[\Delta f_j^{{\bf k}}]^2}{\omega_{{\bf k},j}^2}&\cos(\omega_{{\bf k},j}(t-t'))
 \nonumber \\
 =& \sum_{{\bf k}}\frac{1}{\omega_{{\bf k},j}^2 + \omega_D^2}\cos(\omega_{{\bf k},j}(t-t'))  \nonumber \\
 =&\frac{1}{\omega_D^2}\int_0^\infty\delta(\omega - \omega_{{\bf k},j})
    \frac{\omega_D^2}{\omega^2 + \omega_D^2}\cos(\omega(t-t'))\diff\omega  \nonumber \\
 \approx & \frac{1}{\omega_D^2}\frac{1}{\omega_0}\int_0^\infty\frac{\omega_D^2}{\omega^2 + \omega_D^2}
 \cos(\omega(t-t'))\diff\omega \nonumber \\
 =&\frac{1}{\omega_D^2}\frac{1}{\omega_0}\omega_D \ee^{-\omega_D(t-t')} \nonumber
\end{align}
where $\omega_0\df c/L$.
In this approximation we can then write:
$$
  \bar\gamma_{jm}(\tau) \approx \Theta(t-t')\frac{\gamma B_1^2}{M_s \rho L(1 + \hat h + \hat L') h}
                        \frac{1}{\omega_D^2}\frac{1}{\omega_0}\omega_D \ee^{-\omega_D\tau}.
$$
Integrating this expression between $\tau =0$ and $\tau = \infty$ we find
\begin{align}
 \bar\gamma_{\rm eff} =& \frac{\gamma B_1^2}{M_s \rho L(1 + \hat h + \hat L') h}
                         \frac{1}{\omega_D^2}\frac{1}{\omega_0} \nonumber \\
                      =& \frac{\gamma B_1^2}{M_s \rho L(1 + \hat h + \hat L') h}
		         \frac{h^2}{c^2}\frac{L}{c} \nonumber \\
                      =& \frac{\gamma B_1^2}{M_s \rho (1 + \hat h + \hat L') }
		         \frac{h}{c^3}.
 \label{eq_gamma_eff1}
\end{align}

To be more accurate let's define the functions
$$
\hat g_j(\tau) \df \frac{1}{\hat h(1 + \hat h + \hat L')}\frac{c^2}{L^2}\sum_{{\bf k}}\frac{[\Delta f_j^{{\bf k}}]^2}{\omega_{{\bf k},j}^2}
                     \cos(\omega_{{\bf k},j}(\tau))
$$
so that we can write
$$
\bar\gamma_{jm} = \Theta(\tau)\frac{\gamma B_1^2}{M_s \rho L^2}\frac{L^2}{c^2}\hat g_j(\tau).
$$
The functions $\hat g_j(\tau) $ are plotted in figure 2.
Integrating $\hat g_j(\tau) $ between 0 and $\infty$ we find:
\begin{align}
 \eta\ \df& \int_0^{\infty}\hat g_j(\tau)\diff\tau 
       \approx \frac{h}{c} \nonumber
\end{align}
and then finally
$$
 \bar\gamma_{\rm eff} = \frac{\gamma B_1^2}{M_s \rho c^2}\frac{h}{c},
$$
analogously to what we found previously, \ceq{eq_gamma_eff1}.


\end{document}